\documentclass[preprint,5p]{elsarticle}

\journal{Physics Letters B}

\usepackage{color}
\usepackage{amsfonts}
\usepackage{bm}
\usepackage{hyperref}
\usepackage{caption}
\usepackage{subcaption}
\usepackage{amsmath}
\usepackage{slashed} 

\usepackage{array}
\usepackage{longtable}

\begin{document}

\begin{frontmatter}

\title{Exploring a heavy charged Higgs using jet substructure in a fully hadronic channel}

\author{Riley Patrick\fnref{cor1}}
\ead{riley.patrick@adelaide.edu.au}
\author{Pankaj Sharma\fnref{cor2}}
\ead{pankaj.sharma@adelaide.edu.au}
\author{Anthony G. Williams\fnref{cor3}}
\ead{anthony.williams@adelaide.edu.au}
 \fntext[cor1]{ORCID: 0000-0002-8770-0688}
 \fntext[cor2]{ORCID: 0000-0003-1873-1349}
 \fntext[cor3]{ORCID: 0000-0002-1472-1592}
\address{ARC Center of Excellence for Particle Physics at the Terascale, Department of Physics, University of Adelaide, 5005 Adelaide, South Australia}

\begin{abstract}
In the framework of the type-II Two Higgs Doublet Model (2HDM-II) a charged Higgs search strategy is presented for the dominant production mode $gb \rightarrow tH^\pm$ at the 14 TeV LHC. We consider the decay process which includes $t \rightarrow bW^\pm$ and $H^\pm \rightarrow AW^\pm$, and a fully hadronic final state consisting of $bb\bar{b}+\mbox{jets}+X$. Dictated by the $b \rightarrow s\gamma$ constraints which render $M_{H^\pm} > 480$ GeV we study two scenarios in which the charged Higgs mass is 750 GeV and the pseudoscalar Higgs mass is 200 GeV and 500 GeV. In this mass scheme highly boosted final state objects are expected and handled with jet substructure techniques which also acts to suppress the standard model background. A detailed detector analysis is performed, followed by a multivariate analysis involving many kinematic variables to optimize signal to background significance. Finally the LHC search sensitivities for the two scenarios are presented for various integrated luminosities.
\end{abstract}


\end{frontmatter}


\section{Introduction}

After the breakthrough discovery of a SM-like Higgs particle~\cite{Chatrchyan:2012xdj,Aad:2012tfa}, the Large Hadron Collider (LHC) is expected to unravel the mystery regarding the mechanism of electroweak symmetry breaking (EWSB) during its current and future high energy and high luminosity runs. Although the data corresponding to the newly-discovered particle is consistent with the standard model (SM) predictions for the Higgs production cross sections, decay branching ratios, parity and CP properties, there is still a lot of leeway to probe the extended Higgs sectors beyond the SM (BSM). An obvious and elegant extension of the SM Higgs sector without affecting the $\rho$ parameter is to include another doublet into the theory leading to a two Higgs doublet model (2HDM). A 2HDM enriches the Higgs sector with new scalar states in addition to the light CP even Higgs, $h$, namely, the CP even heavy neutral Higgs, $H$, the CP odd neutral Higgs, $A$, and a pair of charged Higgs, $H^\pm$. Different types of 2HDMs exist in the literature depending on how the two doublets couple with fermions, namely, Type I, Type II, Type Y and Type X. A detailed phenomenology of the various 2HDMs can be found in Ref.~\cite{Branco:2011iw}. 

Any signal of the new scalar states at the LHC would be an unequivocal evidence of new physics (NP). In particular, the search of the charged Higgs has received a significant attention (see Ref. \cite{Akeroyd:2016ymd} and references therein for exhaustive analysis on charged Higgs discovery prospects at the LHC). Of all 2HDMs, in this letter we focus on the  type II 2HDM wherein the $b\to s\gamma$ measurements constrain the charged Higgs mass to be larger than 480 GeV \cite{Misiak:2015xwa}. In this mass regime, we study a heavy charged Higgs production at the LHC in association with a top quark which is the dominant mechanism for $H^\pm$ production when $M_{H^\pm}>M_{\text{top}}$~\cite{Gunion:1986pe}. We further consider the bosonic decays of the charged Higgs to a pseudoscalar via $H^\pm \to W^\pm A$ followed by the decay of pseudoscalar either to a pair of $b$ quarks or via the $Zh$ mode depending upon its mass. The dominant decay of $H^\pm$ in the $tb$ mode suffers from a large $t\bar t$ backgrounds and can have reasonable discovery prospects only in the small or large $\tan\beta$ region~ \cite{Gunion:1993sv,Barger:1993th,Miller:1999bm,Moretti:1999bw,Dev:2014yca}. Recently ATLAS analyzed the Run I data which searched for $H^\pm$ production and decay in the $tb$ mode for charged Higgs mass ranging from 200 GeV to 600 GeV in multijet final states with one electron or muon and found an excess for all charged Higgs mass hypothesis \cite{Khachatryan:2015qxa}. This has stimulated a renewed interest in the search of charged Higgs and motivated a need to reasses the $H^\pm$ search strategies. Other searches for a heavy charged Higgs in type II 2HDM using the top polarization have been performed in Refs.~\cite{Huitu:2010ad,Godbole:2011vw,Rindani:2013mqa,Gong:2014tqa,Gong:2012ru,Cao:2013ud}. 

The bosonic decays of a charged Higgs i.e., $W^\pm h_i$ (where $h_1,h_2,h_3\equiv h,H,A$) open a new domain in the search of $H^\pm$, when they are kinematically accessible. These decays dominate quickly over other modes as soon as they are available \cite{Moretti:2016jkp,Arhrib:2016wpw,Coleppa:2014cca,Li:2016umm}. In Ref.~\cite{Coleppa:2014cca}, the authors have investigated the discovery prospects of $H^\pm$ in $W^\pm H/W^\pm A$ modes with $H/A$ decaying to a pair of $\tau$ leptons and obtained a bound in $M_{H^\pm}$ to 600 GeV with 300 fb$^{-1}$ of data. Ref. \cite{Moretti:2016jkp} discuss the search for $H^\pm$ in $W^\pm b\bar b$ final state originating from the $t\bar b$ and $W^\pm h_i$ decay modes. In Ref. \cite{Li:2016umm}, the authors employed jet substructure techniques to enhance the search prospects in the $W^\pm A$ mode where the heavy charged Higgs leads to highly boosted $W^\pm$ and $A$ bosons assuming a large mass splitting in $H^\pm$ and pseduoscalar. They further made use of multivariate analysis (MVA) namely boosted decision trees (BDT) to optimize the signal significance and demonstrated the utility of substructure techniques in heavy charged Higgs regime.

Given the way the LHC is accruing data in the current run and will continue doing in the future high energy and high luminosity runs, the mass scales of the exotic particles will keep getting pushed to higher values. The decays of such massive particles would lead to a number of boosted daughter particles. When these daughter resonances decay hadronically, the final state in an event would have a large number of ``merged'' jets with a large jet cone radius. This causes a quite difficult situation at the LHC since the existing algorithms and techniques have been focused on the reconstruction of the isolated objects emanated from a slow moving parent resonance. 

To illuminate this rather interesting but experimentally challenging possibility we study the search prospects of $H^\pm$ in a fully hadronic mode at the LHC which leads to a number of merged jets in the final state and employ sophisticated tools including jet substructure techniques and boosted decision trees. In this regard we allow all the SM particles, i.e., the top, the Higgs boson and the $Z/W^\pm$ bosons to decay in hadronic mode. The massive charged Higgs considered in the study would lead to highly boosted SM states in its decay and thus would make the use of jet substructure pertinent. In Ref.~\cite{Li:2016umm} and \cite{Yang:2011jk} the authors have explored the heavy charged Higgs discovery prospects in the $W^\pm A\to W^\pm b \bar b$ and $tb$ decay modes respectively using jet substructure techniques. In probing other Higgs particles in BSM models with extended Higgs sectors jet substructure techniques are utilized in refs.~ ~\cite{Kang:2013rj,Chen:2014dma,Hajer:2015gka,Casolino:2015cza,Conte:2016zjp,Goncalves:2016qhh}.

We organize the paper as follows. The next section discusses the various signal processes and their corresponding backgrounds. Section 3 details the framework of our analysis for two selected benchmark points. We also study signal signifcance using MVA namely the BDT algorithm. Finally we conclude in section 4.

\section{LHC searches for heavy $H^\pm$ and light $A$}
\subsection{Decays of a pseudoscalar Higgs in 2HDM}
\begin{figure}[h!]
\centering
\includegraphics[scale=.6]{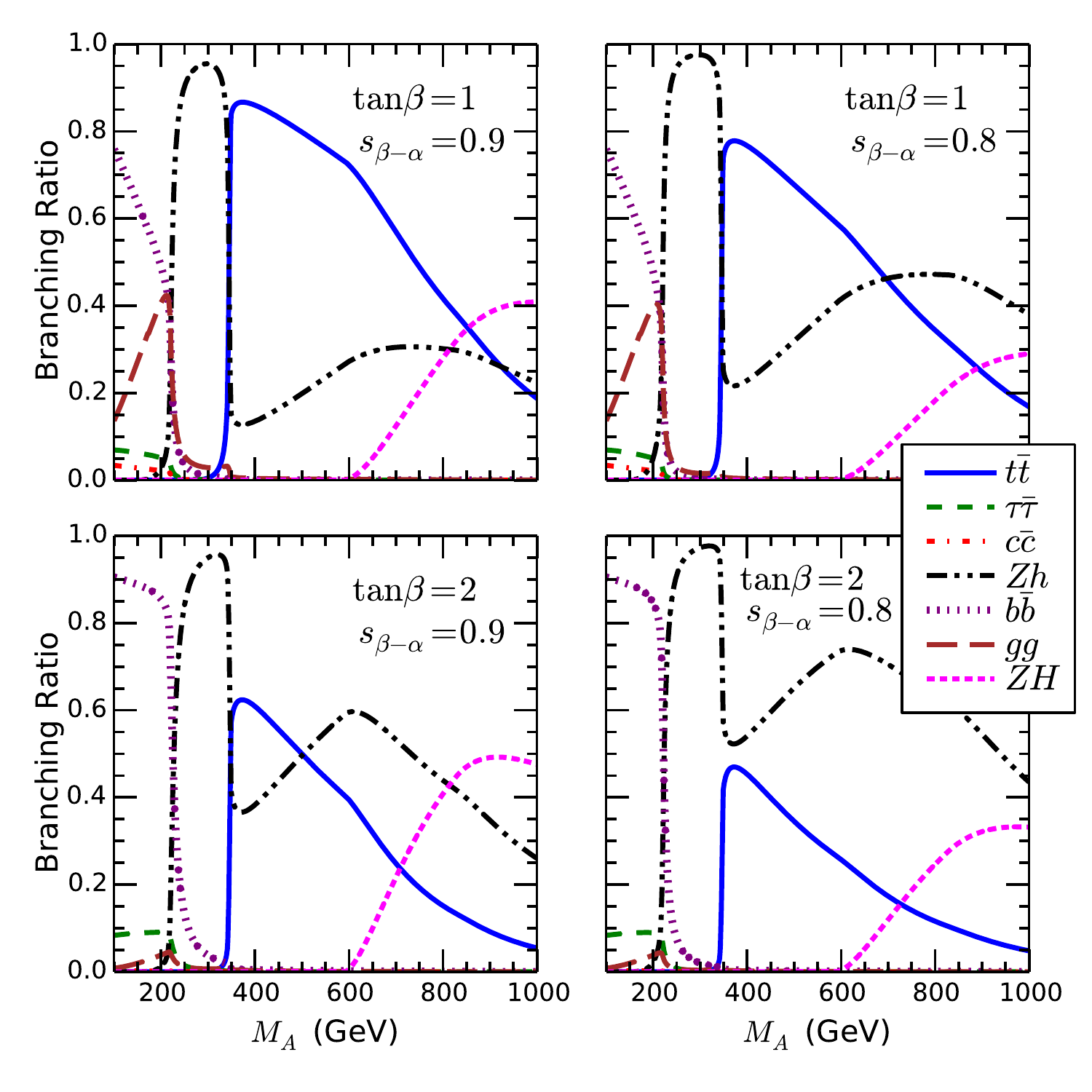}
\caption{\label{fig:br}Branching ratios for the decay of a pseudoscalar for two values of $\tan\beta$ = 1 and 2; and for two values of $\sin(\beta-\alpha)$ = 0.9 and 0.8 in type II 2HDM. The masses for the charged Higgs and the heavy neutral Higgs $H$ are taken to be 750 GeV and 500 GeV respectively.}
\end{figure}

The decays of a heavy charged Higgs have been studied in detail in Ref.~\cite{Li:2016umm}. The conclusion of the analysis is that the bosonic decays of the charged Higgs in $W^\pm A/W^\pm H$ can be dominant in the region when they are kinematically accessible and may become 35-45\% each in certain mass spectra. In this study, we study in detail the decays of a pseduoscalar for two different pseudoscalar masses in type II 2HDM. In Fig. \ref{fig:br} we display the branching ratios (BR) of a pseudoscalar decays to various channels for two different values of $\tan\beta=1$ and 2; and also for two values of $\sin(\beta-\alpha)=0.8$ and $0.9$ starting from pseudoscalar mass from 100 GeV to 1 TeV. The choices of the parameter space are consistent with the current LHC data as has been shown in \cite{Moretti:2016jkp} using tools like \textsc{ HiggsBounds} \cite{Bechtle:2013wla}, \textsc{ HiggsSignals} \cite{Bechtle:2013xfa} and \textsc{ScannerS} \cite{Coimbra:2013qq}. All the decay  branching ratios of $A$ have been obtained using two Higgs doublet model calculater {\tt 2HDMC} \cite{Eriksson:2010zzb}.

\begin{figure*}[htb]
\centering
\includegraphics[scale=.32]{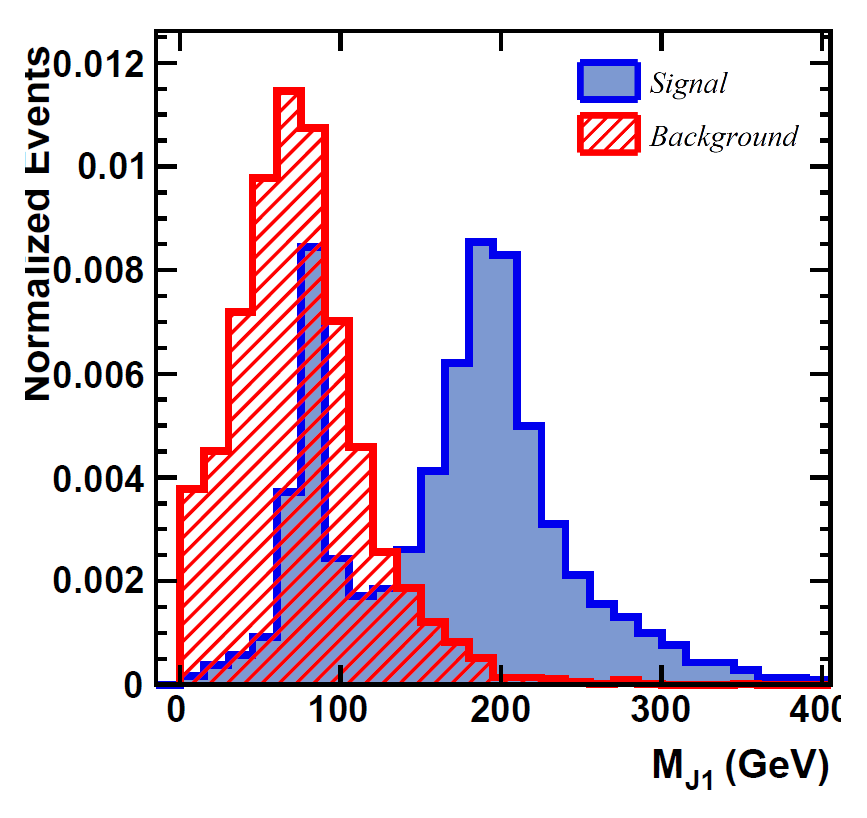}
\includegraphics[scale=.32]{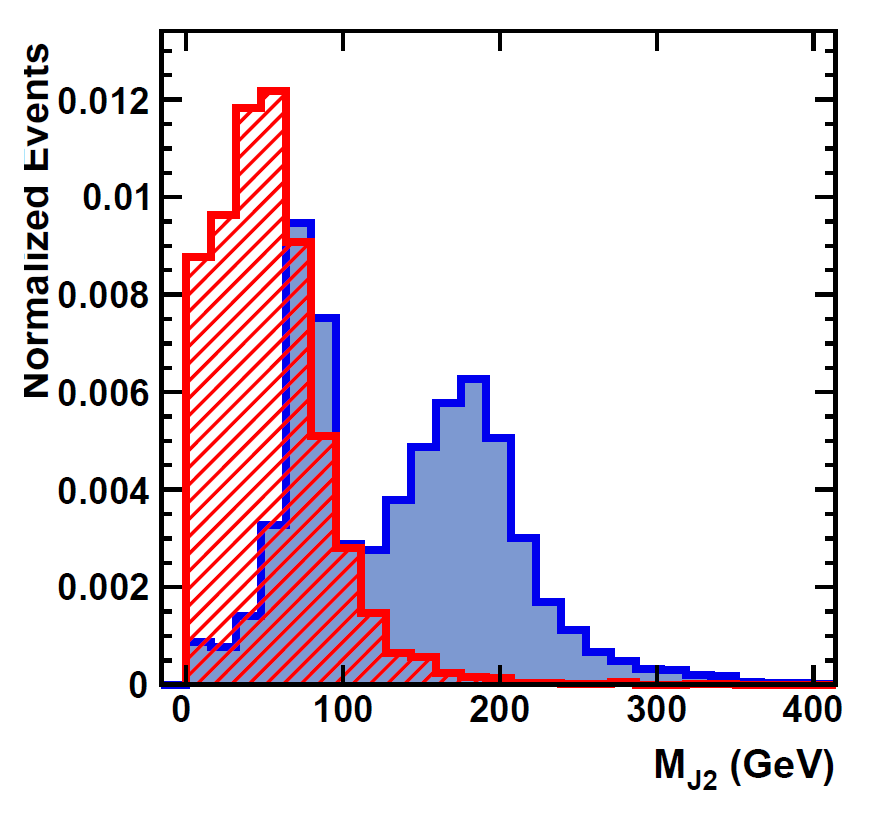}
\caption{\label{fig:b1invmass_jets}Invariant mass distributions of first two leading fat jets for the signal and background in benchmark B1.}
\end{figure*}

From the Fig.~\ref{fig:br}, we notice that the loop induced branching ratio of $A\to gg$ is substantial in the low $\tan\beta$ and $M_A<M_Z+M_h$ region and competes with the $A\to b\bar b$ mode. However in the large $\tan\beta$ region the $gg$ BR goes down rapidly and thereby making the $b\bar b$ mode the only dominant decay channel for $A$. As soon as the $A\to Z h$ channel is kinematically accessible, it becomes quickly dominant with BR$\sim 0.9$. When the $t\bar t$ channel opens up, the BR for the $Zh$ mode is drastically decreased. However, as the mass of pseudoscalar grows and since the $AZh$ coupling is proportional to the momentum transfer in the interaction, the $A\to Zh$ decay restores itself rather quickly and becomes comparable to the $t\bar t$ mode in the low $\tan\beta$ region. However with the large $\tan\beta$, the $t\bar t$ mode becomes suppressed by $1/\tan^2\beta$ and thus the $Zh$ mode becomes the dominant decay mode for a pseudoscalar. 

In this study, we consider two different values for the pseudoscalar mass, i.e., 200 GeV and 500 GeV in order to exhibit the distinct signal characteristics. For a 200 GeV pseudoscalar $A$, we study the $b\bar b$ decay mode while for a 500 GeV $A$ the $Zh$ mode is studied. One could argue that for an approximately 300 GeV pseudoscalar, the $Zh$ branching ratio is the largest, despite this we choose the 500 GeV mass as the decay products, $Z$ and $h$, are highly boosted and thus is the most suitable benchmark point for the jet substructure analysis.

\subsection{Signature and Backgrounds}
We study a charged Higgs produced in association with a top quark at the 14 TeV LHC followed by the decay $H^\pm \rightarrow W^\pm A$ in the heavy charged Higgs mass scenario. As previously mentioned, we consider two masses for the pseudoscalar, i.e., 200 GeV and 500 GeV which lead to two distinct signal kinematics. For a 200 GeV pseudoscalar, the $A\to b\bar b$ deay is considered. For this signal, we consider an entirely hadronic channel, including 3 $b$ jets from the $A$ and the top decays, whilst 4 light jets are present from the $W^\pm$ decay, leading to the signal $bb\bar{b}jjjj+X$. The background with highest cross section is $WWbbj$, which includes top pair production plus one jet process, as the light jet may be mistagged as a $b$ jet in a non-negligible number of events. The irreducible background is the $WWbbb$ process which comes from top pair production in association with a $b$ jet. Finally the background from the $WWbjj$ process is a manageable background despite its high cross section due to $b$ tagging.

\begin{figure*}[htb]
\centering
\includegraphics[scale=.25]{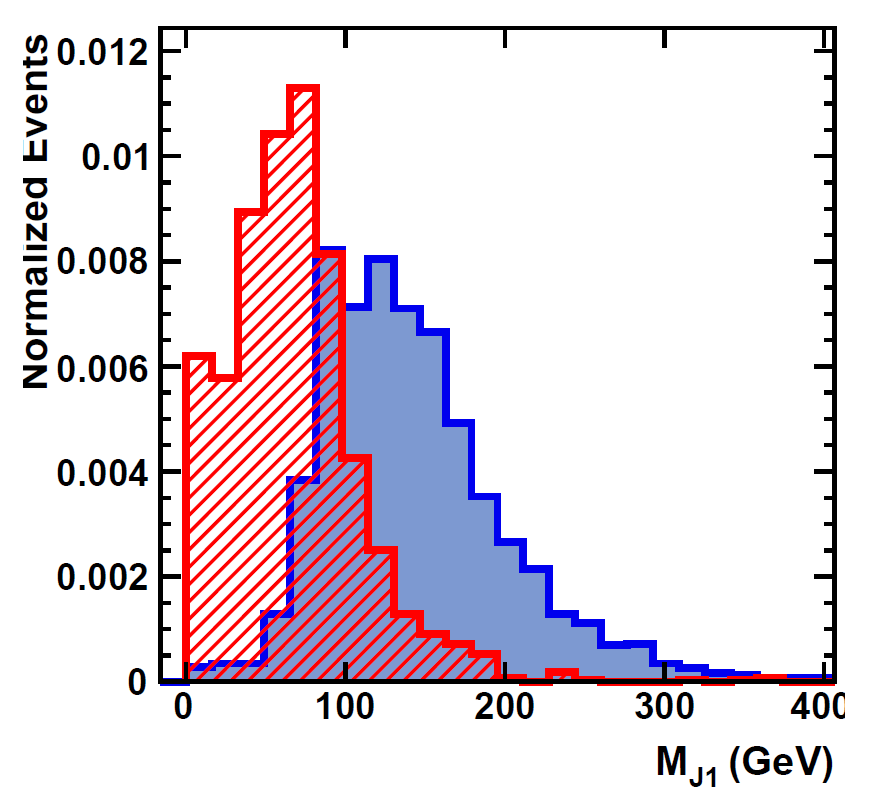}
\includegraphics[scale=.25]{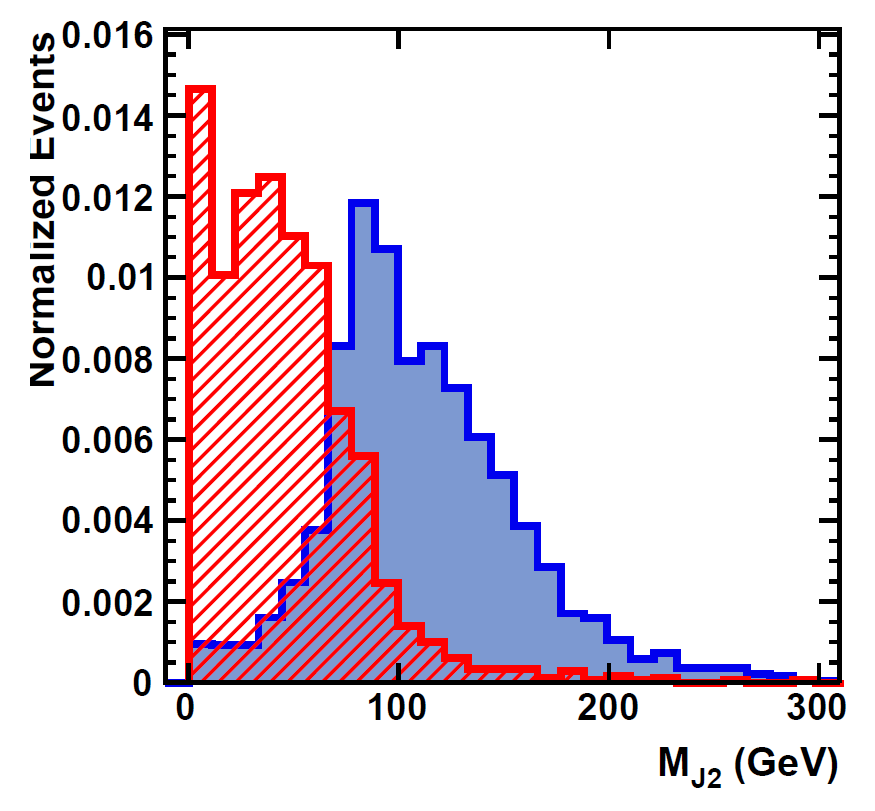}
\includegraphics[scale=.25]{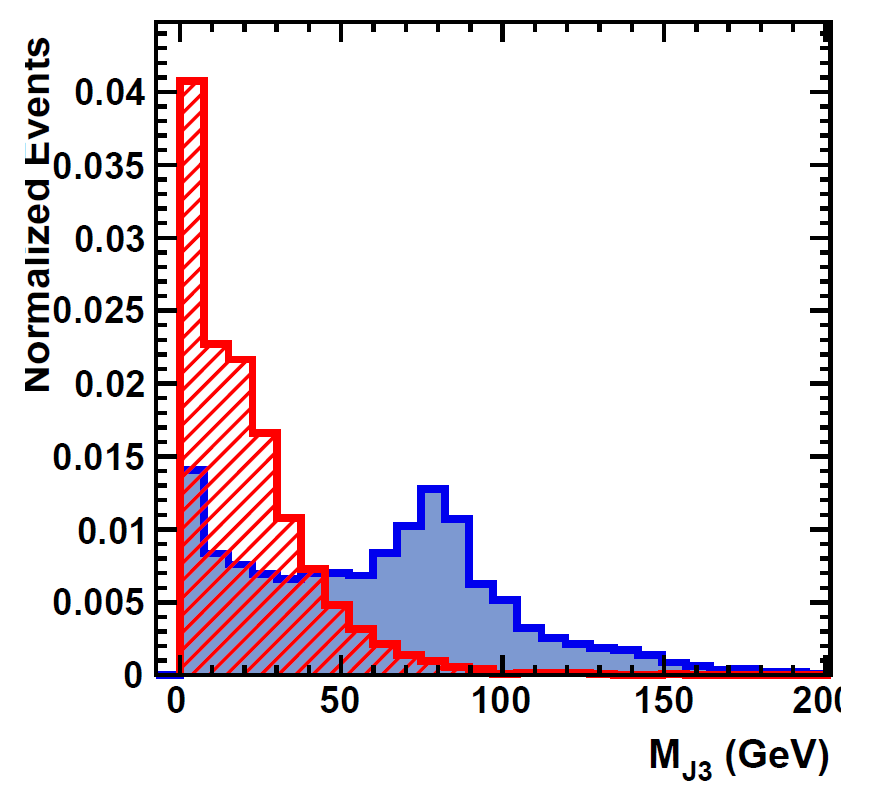}
\caption{\label{fig:b2invmass_jets} Invariant mass distributions of first three leading fat jets for the signal and background in benchmark B2.}
\end{figure*}

For the CP odd Higgs boson mass of 500 GeV, we consider it to be decaying via the $Zh$ mode. As the pseudoscalar in this scenario is heavy, it leads to boosted $Z$ and $h$ bosons. Thus in this signal region, we expect to find an additional fat jet with substructure corresponding to a $Z$ boson. The dominant background considered for this signal in our analysis is $t\bar t+3j$. 

The signal and background events were generated at leading order with {\tt Madgraph5} \cite{Alwall:2014hca} with proper matching of the jets with parton shower using the CKKW algorithm. The events are then passed to \texttt {PYTHIA8.2} \cite{Sjostrand:2007gs} to perform parton showers and hadronization. All events were then passed to {\tt DELPHES3} \cite{deFavereau:2013fsa} for the fast detector simulation, where we apply the default  ATLAS detector card. The \texttt {DELPHES3} output is then used for jet substructure analysis using \texttt {FastJet}\cite{Cacciari:2011ma}.

\subsection{Benchmark points for the analysis}

The motivation of the benchmark point choice for this study was two-fold. Firstly, it is known that jet substructure analysis performs optimally in environments when jets are highly boosted and collimated (and thus form merged cone of large radius). This means that the mass difference between the charged Higgs and the pseudoscalar Higgs ought to be high to ensure a large boost of this psuedoscalar Higgs. Secondly, it was noted in Ref.~\cite{Li:2016umm}, which uses the same parameter space as this study, that a 750GeV charged Higgs provides a close to optimal balance of cross section and jet substructure utility. That is to say, as the charged Higgs mass rises the production cross section goes down, but the power of jet substructure methods increases - these two effects appear to give the 750 GeV charged Higgs the best discovery potential. 
\begin{figure*}[htb]
\centering
\includegraphics[scale=.34]{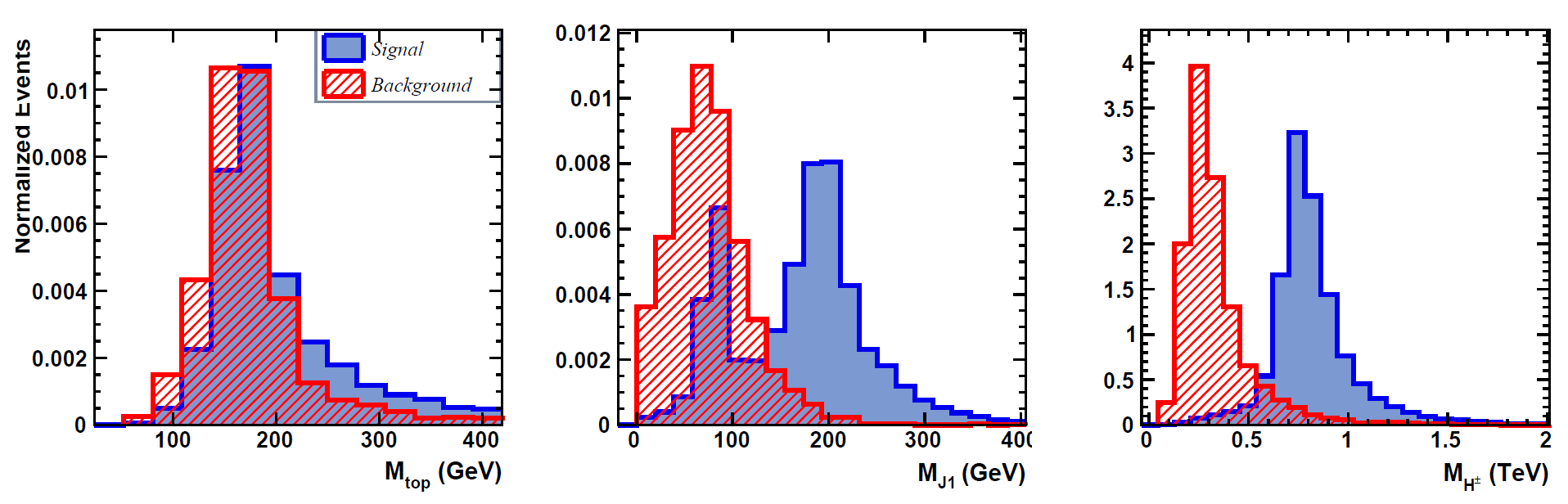}
\includegraphics[scale=.34]{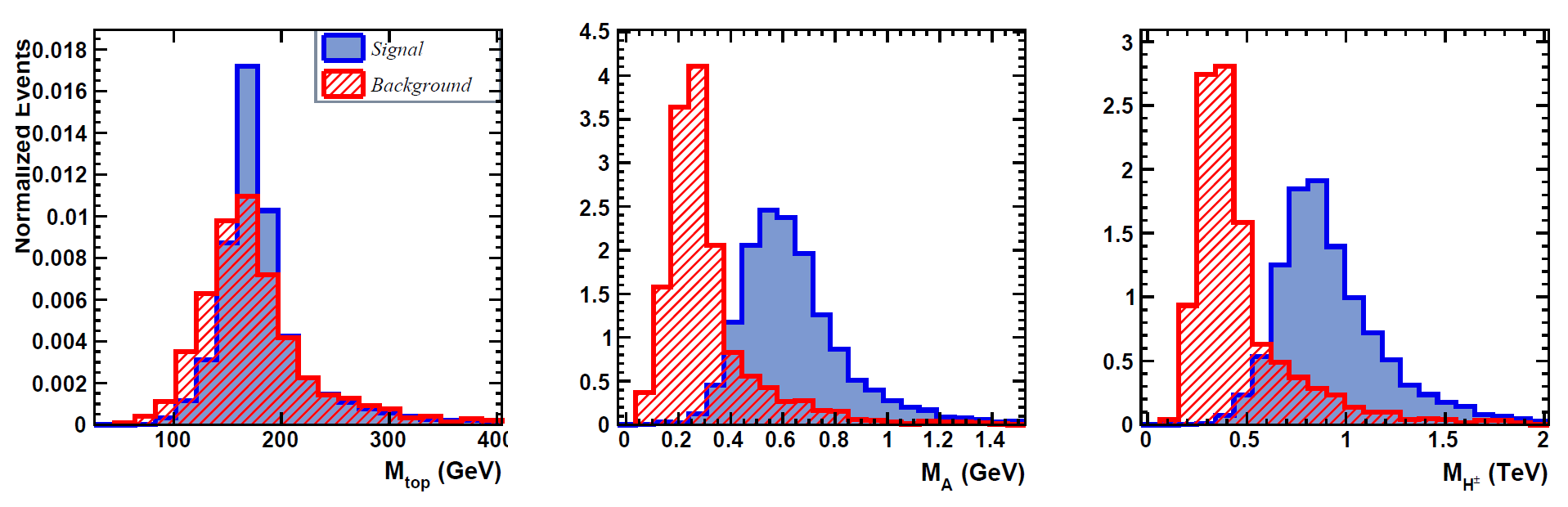}
\caption{\label{fig:invmass}Invariant mass distributions of reconstructed variables for the signal and background in B1 (top row) and B2 (bottom row).}
\end{figure*}
With this reasoning two benchmark points are chosen with a fixed charged Higgs mass of 750 GeV. Benchmark 1 (B1) has a pseudo-scalar mass of 200 GeV wherein the dominant decay process of interest is $A\rightarrow b \bar{b}$ which will give a fat jet that will have two subjets which may be b-tagged. Benchmark 2 (B2) has a pseudo-scalar mass of 500 GeV wherein the dominant decay of interest is $A\rightarrow Zh$. B2 provides an extra opportunity to apply jet substructure methods as the pseudoscalar Higgs now contributes an extra fat jet to the final state due to $Z\rightarrow jj$ and $h\rightarrow b \bar{b}$. We still search for a fat jet with twice b-tagged substructure, however there is a richer set of kinematics in the final state due to this extra fat jet. 

$\newline$In B1 we take $M_h=125$ GeV, $\tan\beta=1$ and $\sin(\beta-\alpha)=0.9$. As the analysis for this benchmark point does not depend on the CP property of the neutral scalars that $H^\pm$ decays to, it is equally applicable to signals in which $H^\pm\to W^\pm h/H$. As mentioned in the foregoing, the current LHC data prefers the alignment scenario leading to almost equal coupling of the pseudoscalar and the heavy CP even scalar to charged Higgs. Thus including the contribution of $H$ into our analysis may further improve the signal cross section and in turn achieve a better signal-to-background ratio. In B2 we take $M_h=125$ GeV, $\tan\beta=2$ and $\sin(\beta-\alpha)=0.8$. 
We also choose $M_H=500$ GeV, degenerate with the pseudo-scalar, effectively setting Br($A\rightarrow ZH$)$=0$. Note that this decay, followed by $H\rightarrow WW/ZZ$ when $A\to ZH$ is allowed, may be analysed in the same way, but with an even richer set of final state kinematics due to the presence of even more jets.

\section{Analysis}
\subsection{Framework}
First we demand $p_T^{j,b}>20~\text{GeV},~ |\eta_{j,b}|<2.5$. Next we preselect relevant physics objects - the particle-flow charged tracks (after isolating the charged leptons), the particle-flow neutral hadrons, and the particle-flow photons from the {\tt DELPHES3} output for using in jet reconstruction. These objects are clustered with the Cambridge-Aachen (CA) jet algorithm \cite{Dokshitzer:1997in} with a particular cone size of $R = 1.2$ in order to capture all collimated decay products of the $W^\pm$ and $A$ bosons, and they are called ``fat jets''. The Butterworh-Davison-Rubin-Salam (BDRS) algorithm \cite{Butterworth:2008iy} is then applied, utilizing the Mass-Drop (MD) tagger, to identify substructure in the clustered jets earlier identified as fat jets. The fat jets are then filtered by reclustering the constituents with radius $R_{\text{filt}} = \text{min}(0.35;R_{12}/2)$ and selecting the three hardest subjets to suppress pile-up effects. A Higgs tagger is then employed to discover which fat jet corresponds to the Higgs. This amounts to demanding that two of the three subjets be $b$ tagged - we take 70\% $b$ tagging efficiency and 1\% mis-tagging rate for light flavor and gluon jets. After the Higgs, and all other fat jets are successfully identified we sideline all constituents of these jets, and re-cluster all remaining objects with cone radius $R = 0.4$ and call them ``narrow jets''.

\subsubsection{Benchmark 1}

After the jet substructure analysis has been performed in B1 we expect at least two fat jets, one of which is tagged as a Higgs, corresponding to the $W^\pm$ and $A$ decaying from the charged Higgs. In a small number of events we may also see a third fat jet, corresponding to the decay products of the top quark. 

It is clear from the peaks at 200 GeV in Fig.~\ref{fig:b1invmass_jets} that the pseudoscalar Higgs can be identified as one of the two hardest fat jets in a large percentage of events, whilst the other of these fat jets can be identified as a $W$ boson. Motivated by this the charged Higgs is reconstructed from the Higgs jet and the hardest remaining fat jet.  Top reconstruction is more involved - first a $W$ boson is reconstructed by calculating the invariant mass of every combination of two narrow jets and choosing the combination which minimizes $|M_W-M_{j1j2}|$. The top is then reconstructed from this $W$ and the hardest remaining narrow jet. This reconstruction strategy for B1 appears to be effective as shown in Fig.~\ref{fig:invmass}.

\subsubsection{Benchmark 2}
In B2 we expect to see at least three fat jets in the final state corresponding to the $Z$, $W$ and $h$, and once again we may see in a small number of events a fat jet corresponding to the top quarks decay. We see in top panel of Fig.~\ref{fig:b2invmass_jets} that the three hardest fat jets appear to identify as the $Z$, $W$ and $h$ - these invariant mass distributions have peaks which correspond to the expected masses. The leading fat jet has a very wide peak due to the merging of three distributions peaked at the masses of these three particles. The second hardest fat jet also appears to have this property but with a more dominant peak at the $Z$ and $W$ masses, implying the $h$ is most often the hardest fat jet. Finally the softest of the three fat jets has a peak over the $W$ and $Z$ masses, but little to no features above these masses, implying this jet is most often the $W$ decaying directly from the charged Higgs. Also of note is that for background distributions, only the hardest fat jet appears to have a clear peak at the $W$ mass, whilst the second hardest fat jet appears to only have a bump on the right side of the distribution, and finally the third fat jet appears to have no significant features at all. This indicates that in the majority of background events only one fat jet, the hardest one, is correctly attributed to the top decay products, whilst the two softer fat jets are soft radiation lying inside a small cone radius.

Motivated by the above and also noting that the decay products of the pseudoscalar Higgs will have in general a greater boost than the $W^\pm$ we reconstruct it from the Higgs jet and the hardest remaining fat jet. The charged Higgs is then reconstructed from the pseudoscalar Higgs and the hardest remaining fat jet. The top is dealt with in the same way as for B1. Once again, the reconstruction appears to work very well as can be seen in bottom panel of Fig.~\ref{fig:invmass}.

\subsection{Multivariate Analysis}
A cut based analysis was found to be insufficient in many scenarios similar to that of this study, that is, for a charged Higgs mass in the range of 500 to 1000 GeV \cite{Li:2016umm}. That being said, a multivariate analysis will provide as good, if not better, signal discrimination as a cut-based analysis as long as the necessary variables are included in the analysis. Motivated by this, multivariate analysis techniques are applied to optimize the signal to background discrimination, namely the Boosted Decision Tree (BDT) algorithm is implemented. This algorithm is chosen as it has been shown to provide extremely accurate event classification in a variety of high energy physics scenarios~\cite{Li:2016umm,Yang:2011jk,Gallicchio:2010dq,Sahin:2016qgg}. Other methods such as Multi-Layered Perceptron (MLP) in artificial neural network were used in preliminary studies and provided excellent classification accuracy but were deemed too slow (when compared to BDT). The BDT algorithm was implemented within the {\tt TOOLKIT FOR MULTIVARIATE DATA ANALYSIS WITH ROOT} ({\tt TMVA}) \cite{root:tmva} which provides a flexible and simple to use framework for classification and regression problems in high energy physics.

\begin{figure}[!h]
\centering
\includegraphics[scale=.4]{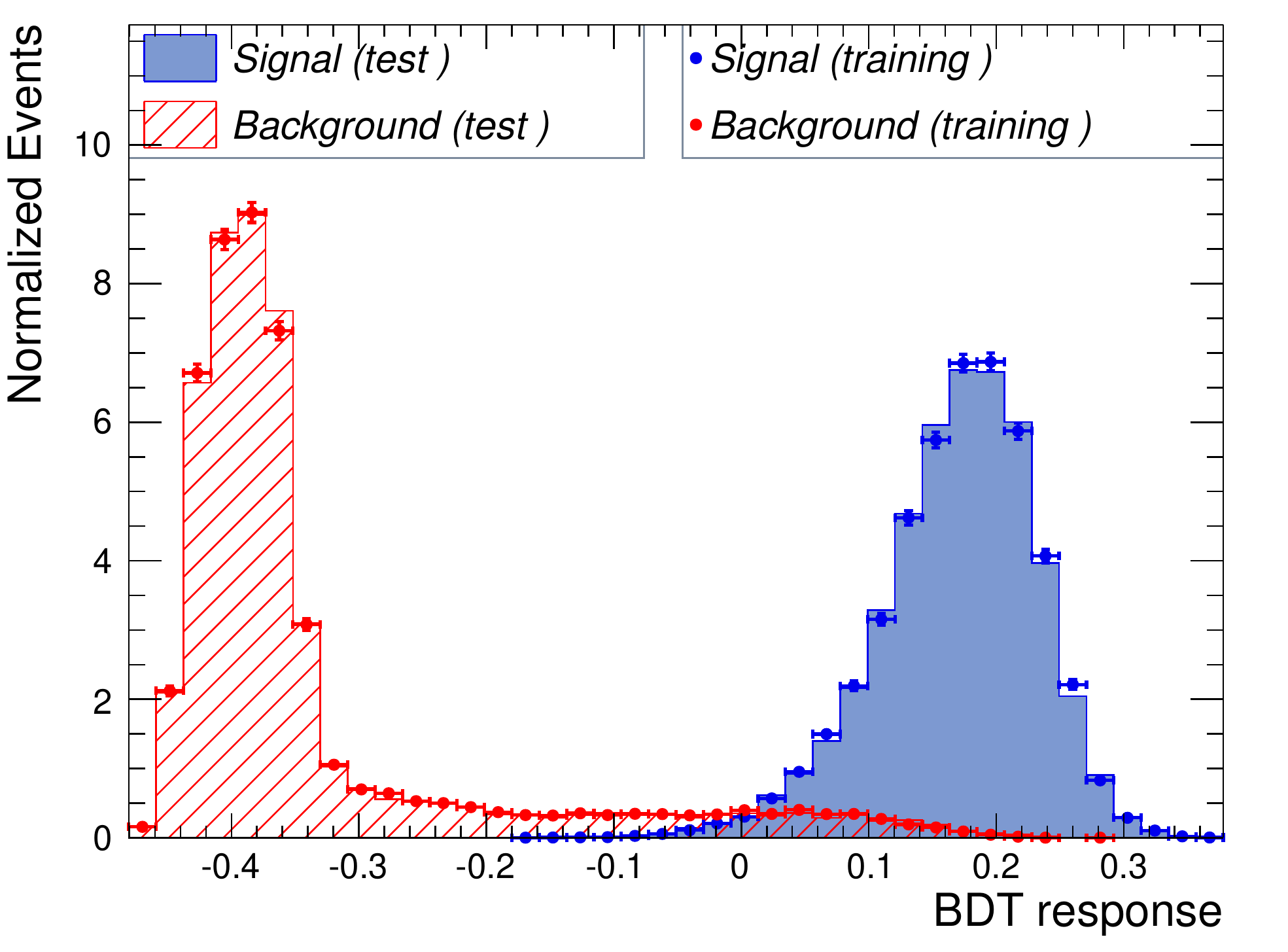}\\\vspace{0.25cm}
\includegraphics[scale=.4]{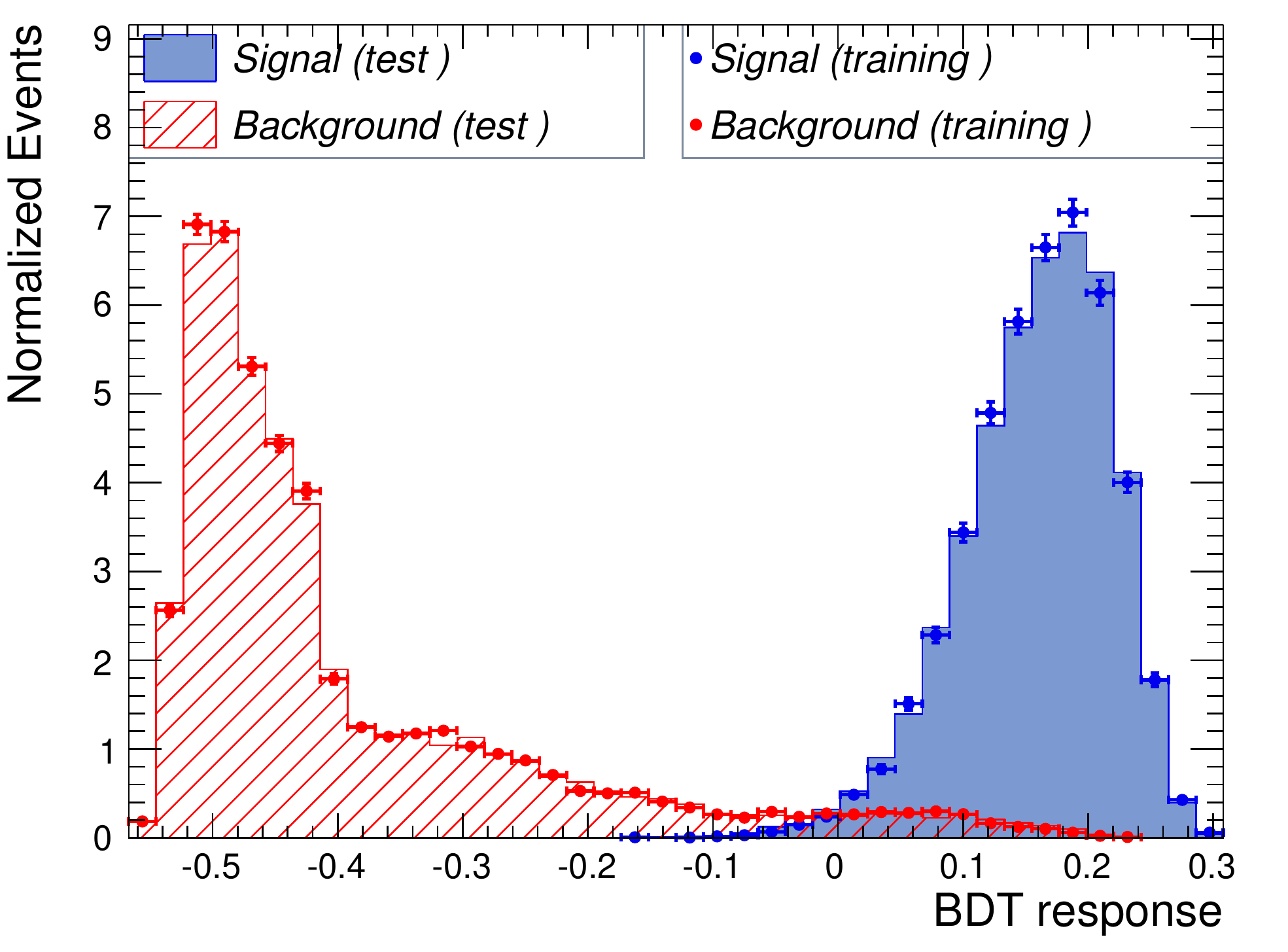}
\caption{\label{fig:bdt_dist} BDT variable distribution for B1 (top panel) and B2 (bottom panel).}
\end{figure}

\begin{figure}[!h]
\centering
\includegraphics[scale=.4]{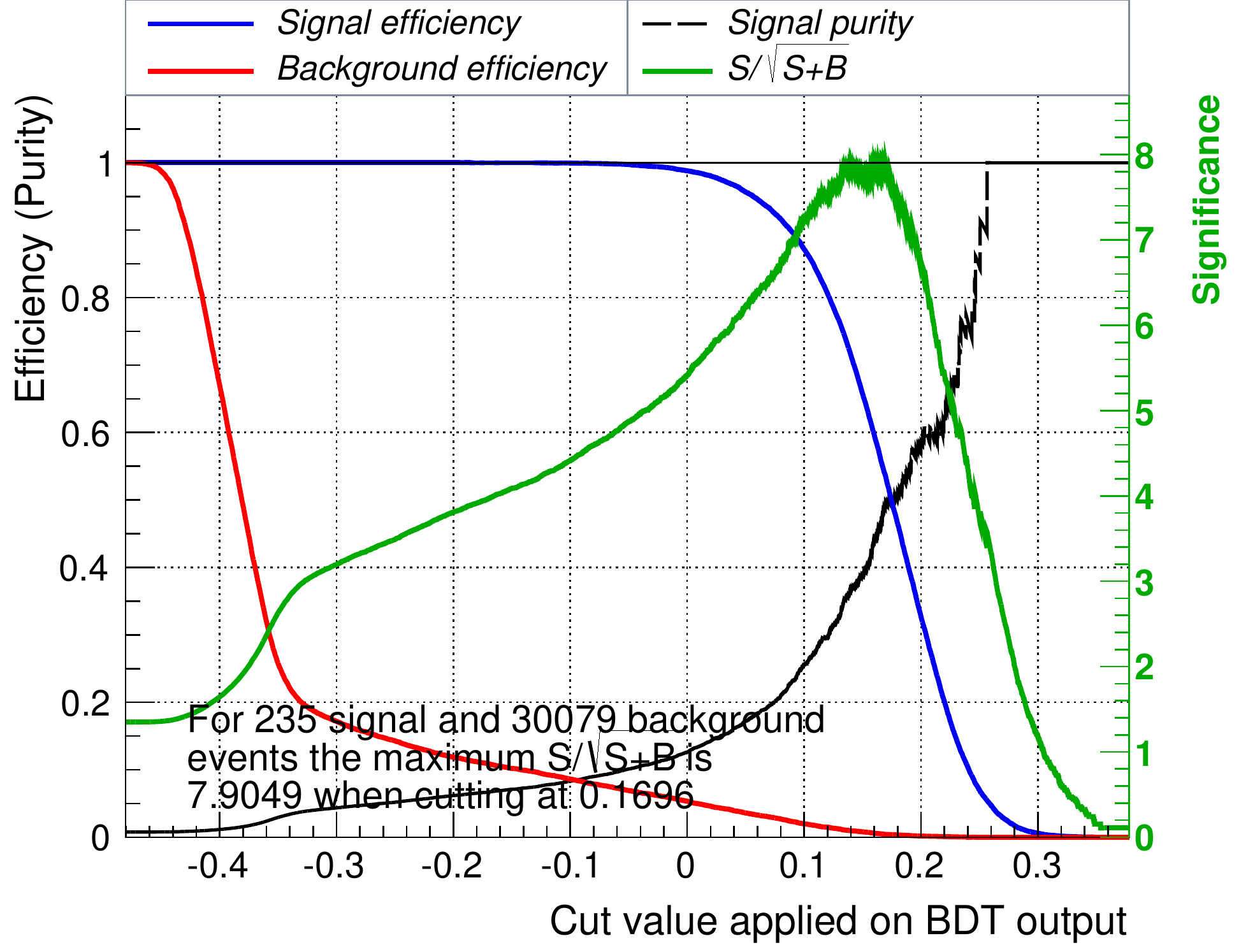}\\\vspace{0.5cm}
\includegraphics[scale=.4]{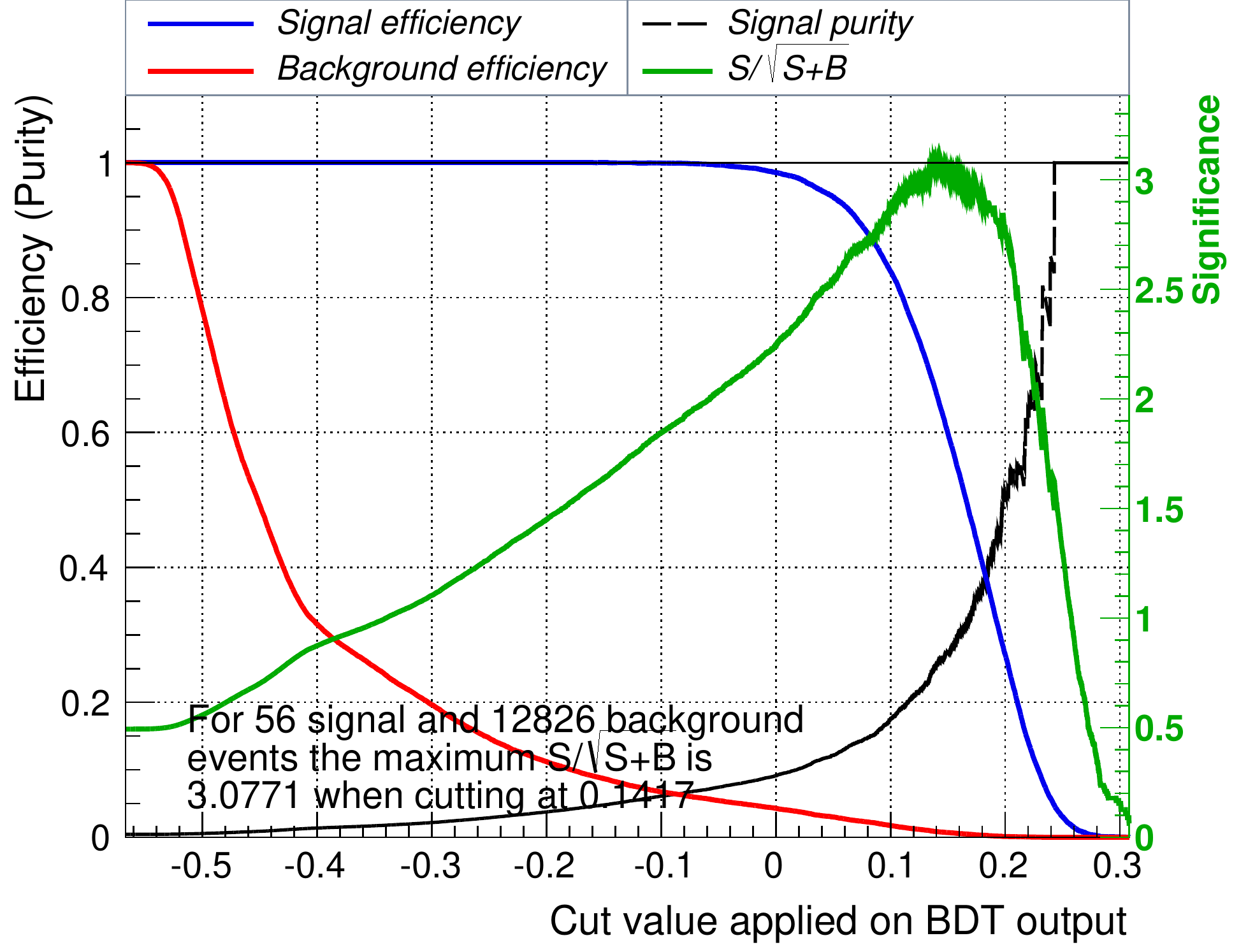}
\caption{\label{fig:significance} Significance and purity distribution curves for B1 (top panel) and B2 (bottom panel).}
\end{figure}

When choosing the variables that will be optimized over it is of greater importance to ensure variables with high discriminating power are included than it is to ensure variables with low discriminating power are excluded. This is because BDT algorithms are rather robust and resistant to overtraining when steps to avoid it are taken\cite{root:tmva}. Most importantly independent training and evaluation datasets are used, as well as boosting and pruning. For B1 and B2 there were nine and thirteen variables chosen comprised of: 
\begin{itemize}
\item Transverse momentum of fat jets
\item Invariant mass of fat jets
\item Transverse momentum of reconstructed particles
\item Invariant mass of reconstructed particles
\item Scalar sum of transverse momentum
\end{itemize} 
Prior to passing these variables through the BDT algorithm, we must define our pre-selection and include their effect on our significance. We choose the aforementioned trigger cuts - $p_T^{j,b}>20~\text{GeV},~ |\eta_{j,b}|<2.5$ and the presence of two (B1) and three (B2) fat jets, of which one is tagged as a Higgs jet. As well as this we also demand that in BP1 that $P_T^{F1} > 260$ GeV and$P_T^{F2} > 200$ GeV, while in B2 we demand that $P_T^{F1} > 200$ GeV, $P_T^{F2} > 160$ GeV and $P_T^{F3} > 100$ GeV. The requirements of two and three fat jets (one of which must be tagged as a Higgs) in B1 and B2 respectively prove to be quite detrimental to backgrounds in respective cases. The resulting pre-selection efficiencies, including b-tagging/mis-tagging efficiency, can be seen in Table~\ref{tab:presel_eff}. As an indication of BDT performance Fig.~\ref{fig:bdt_dist} presents the BDT response distributions for the signal and background for training and test data - the training and test distributions are very similar in all cases indicating that overtraining has been avoided. In Fig. \ref{fig:significance} we plot the significance and purity curves, defined as $S/\sqrt{S+B}$ and $S/(S+B)$ respectively for benchmark points B1 (top panel) and B2 (bottom panel) - which give an indication of the discrimination of signal against background, whilst the former is taken as the final indicator of signal significance.

\begin{table}[h!]
\begin{center}
\newcolumntype{C}[1]{>{\centering\let\newline\\\arraybackslash\hspace{0pt}}m{#1}} 
\begin{tabular}{ ||C{2.5cm}||C{2.5cm}| C{2.5cm} ||}
\hline\hline
\textbf{Benchmark Points} & \textbf{Signal Efficiency (\%)} & \textbf{Background Efficiency (\%)} \\\hline\hline
BP1 & 15.6 & $\approx$~6.6$\times$10$^{-2}$	\\\hline
BP2 & 20.2 & $\approx$~5.21$\times$10$^{-2}$	 \\\hline
\hline
\end{tabular}
\caption{ Signal significance defined as $S/\sqrt{S+B}$ of B1 and B2 for three different values of integrated luminosities at the 14 TeV LHC. \label{tab:presel_eff}}
\end{center}
\end{table}

Considering the pre-selection efficiencies, at 100 fb$^{-1}$ of integrated luminosity the number of expected signal and background events for B1 and B2 are 235 and 30079, and 56 and 12826, respectively. The signal significance for three different values of integrated luminosities, i.e., 100 fb$^{-1}$, 500 fb$^{-1}$ and 1000 fb$^{-1}$ resulting from the multivariate analysis can be seen in Table~\ref{tab:sigtable}.
\begin{table}[h!]
\begin{center}
\newcolumntype{C}[1]{>{\centering\let\newline\\\arraybackslash\hspace{0pt}}m{#1}} 
\begin{tabular}{ ||C{2.25cm}||C{1.5cm}| C{1.5cm} | C{1.5cm} ||}\hline \hline
&\multicolumn{3}{c||}{\textbf{Signal Significance}}    \\\cline{1-4}
\textbf{Benchmark Points} & 100~fb$^{-1}$ & 500~fb$^{-1}$ & 1000~fb$^{-1}$ \\\hline
BP1 & 3.7 & 8.2 & 11.7	\\\hline
BP2 & 3.1 & 6.9 & 9.8	 \\\hline
\hline
\end{tabular}
\caption{ Signal significance defined as $S/\sqrt{S+B}$ of B1 and B2 for three different values of integrated luminosities at the 14 TeV LHC. \label{tab:sigtable}}
\end{center}
\end{table}

Despite the overwhelming number of background events and relatively small number of signal events the BDT performs quite well. This is demonstrated by the fact that B1 and B2 can be probed as early as approximately 250 fb$^{-1}$ of integrated luminosity which is quite significant given that it is the most challenging scenario considered in the study. 

Another interesting scenario for benchmark point B2 is when $h\to gg$ and $h\to c\bar c$ for $\tan\beta=1$ and $\sin(\beta-\alpha)=0.9$ which together constitute 45\% of total $h$ decays. This results in highly boosted and collimated light jets merging together to form a ``Higgs jet''. Although the cross section for this scenario can be as large as it is for B2 in this study, the lack of $b$ tagging would result in considerably more background events. However as noticed in our analysis, the requirement of three fat jets is sufficient to suppress the background and $b$ tagging affects both signal and background events in a similar manner, thus the $h\to jj$ can also be an interesting mode to probe B2 at the LHC. Also realizing that the event kinematics would remain the same for the $h\to b\bar b$ and $h\to j j$ decay modes, a naive estimate for the significance has been obtained and it is found that the signal significance for the $h\to jj$ mode would be close to that of the $h\to b\bar b$ decays.

\section{Summary \& Conclusion}
Since the discovery of the 125 GeV scalar particle currently identified as the Standard Model Higgs at the Large Hadron Collider a new era of experiment around the mechanism of Electro-Weak Symmetry Breaking has begun. Any discovery of a new scalar would be a clear mark of physics beyond the standard model.

This letter has explored fully hadronic decays of the charged Higgs within the framework of the Type-II Two Higgs Doublet Model (2HDM-II). Specifically we have looked at two similar, but distinct, processes - both beginning with charged Higgs production in association with a top quark which is followed by the decay $H^{\pm}\rightarrow WA$. The first benchmark includes $A\rightarrow b\bar{b}$, whilst the top and $W$ decay fully hadronically - in this scenario we choose $M_A=200$. The second benchmark includes $A\rightarrow Zh$, where $Z$, $h$, $W$ and top all decay fully hadronically - in this scenario we choose $M_A=500$. In both scenarios the charged Higgs mass is chosen to be 750 GeV. These mass choices result in highly boosted final state objects.

To enhance the search prospects we utilize jet substructure methods, and Higgs tagging of the pseudoscalar Higgs. A detailed detector analysis is performed on both benchmark points and a multivariate analysis is also performed to optimize the signal to background significance. Specifically, we utilize the Boosted Decision Tree algorithm which takes a variety of kinematic variables as input and outputs a single multi-dimensional variable and a cut on this variable which will produce the optimal signal to background ratio. The results of this multivariate analysis indicate that both benchmark points are discoverable with approximately 100 fb$^{-1}$ of luminosity.

As the LHC is rapidly accumulating data and moving the scale of new particles higher and higher, the role of jet substructure techniques becomes more pertinent for searches of new resonances. Stimulated by this fact, we demonstrate the utility of sophisticated and state-of-the-art techniques like jet substructure and multivariate analysis in a highly experimentally challenging scenario of full hadronic final states in the search of a heavy charged Higgs in type-II 2HDM. Though in our analysis we consider type II 2HDM, the results can be easily applied to other 2HDMs by normalizing with cross section and branching ratios in respective models.

\section*{Acknowledgments}
This work is supported by the University of Adelaide and the Australian Research Council through the ARC Center of Excellence for Particle Physics (CoEPP) at the Terascale (grant no. \ CE110001004).

\end{document}